\documentclass[reprint,pre,showpacs, bibnotes, footinbib]{revtex4-1}
\pdfoutput=1
\usepackage{graphicx}
\usepackage{graphicx,amsfonts}
\usepackage{amsmath}
\usepackage{verbatim}

\usepackage{natbib}
\usepackage{color}
\definecolor{darkblue}{rgb}{0,0,0.5}
\definecolor{darkred}{rgb}{0.5,0,0}
\usepackage[colorlinks=true,urlcolor=darkblue,citecolor=darkblue,linkcolor=darkred,hyperfootnotes=false]{hyperref}

\begin{document}

\newcommand{\atanh}
{\operatorname{atanh}}
\newcommand{\ArcTan}
{\operatorname{ArcTan}}
\newcommand{\ArcCoth}
{\operatorname{ArcCoth}}
\newcommand{\Erf}
{\operatorname{Erf}}
\newcommand{\Erfi}
{\operatorname{Erfi}}
\newcommand{\Ei}
{\operatorname{Ei}}

\newcommand{\qtot}{Q_{\tt tot}}
\newcommand{\qdtot}{Q^\dagger_{\tt tot}}
\newcommand{\qex}{Q_{\tt ex}}
\newcommand{\qhk}{Q_{\tt hk}}
\newcommand{\sigminf}{\Sigma_\infty}
\newcommand{\fqtot}{\mathcal{Q}_{\tt tot}}
\newcommand{\fqdtot}{\mathcal{Q}^\dagger_{\tt tot}}
\newcommand{\fqex}{\mathcal{Q}_{\tt ex}}
\newcommand{\fqhk}{\mathcal{Q}_{\tt hk}}
\newcommand{\F}{\mathcal{F}}
\newcommand{\Pp}{\mathcal{P}}
\newcommand{\rhoss}{\rho_{\tt SS}}
\newcommand{\Yhasa}{\mathcal{Y}}
\newcommand{\tYhasa}{\tilde{\mathcal{Y}}}
\newcommand{\J}{\mathcal{J}}
\newcommand{\C}{\mathcal{C}}
\newcommand{\xvec}{\vec{x}}
\newcommand{\evec}{\vec{e}}
\newcommand{\avec}{\vec{\alpha}}
\newcommand{\tvec}{\vec{\theta}}
\newcommand{\lvec}{\vec{\lambda}}
\newcommand{\R}{\text{R}}
\newcommand{\hk}{\text{hk}}
\newcommand{\tot}{\text{tot}}
\newcommand{\Ss}{\mathcal S_{\tt s}}
\newcommand{\ex}{\text{exp}}
\newcommand{\Sna}{\mathcal{S}_{\tt na}}
\newcommand{\Ana}{\mathcal{A}_{\tt na}}
\newcommand{\Snal}{S_{\tt na}}
\newcommand{\Anal}{A_{\tt na}}
\newcommand{\dAna}{\dot{\mathcal{A}}_{\tt na}}
\newcommand{\Sex}{\mathcal{S}_{\tt ex}}
\newcommand{\Aex}{\mathcal{A}_{\tt ex}}
\newcommand{\Sexin}{S_{\tt ex}}
\newcommand{\Aexin}{A_{\tt ex}}
\newcommand{\pt}{p_m}
\newcommand{\ps}{p^{\tt s}}
\newcommand{\psj}{p_{m_j}^{\tt s}(\sigma_{t_j})}
\newcommand{\psjone}{p_{m_{j-1}}^{\tt s}(\sigma_{t_j})}
\newcommand{\pin}{p_0}
\newcommand{\pfin}{p_{m_N}(t)}
\newcommand{\hP}{\hat{P}_{\tt s}}
\newcommand{\htP}{\hat{\tilde{P}}}
\newcommand{\hY}{\hat{Y}}
\newcommand{\hM}{\hat{M}}
\newcommand{\hMM}{\hat{\mathbb{M}}}
\newcommand{\e}{\text{e}}
\newcommand{\s}{\text{s}}
\newcommand{\ptwo}{p^{(2)}}
\newcommand{\mL}{\hat{\mathcal{L}}}
\newcommand{\mK}{\mathcal{K}}
\newcommand{\refer}{{\tt ref}}
\newcommand{\Tef}{T_{\tt eff}}
\newcommand{\gef}{\gamma_{\tt eff}}
\title{Non-adiabatic entropy production for non-Markov dynamics}
\author{Reinaldo Garc\'ia-Garc\'ia}
\email{reinaldo.garcia@cab.cnea.gov.ar}
\affiliation{Centro At\'omico Bariloche and Instituto Balseiro, 8400 S. C. de Bariloche, Argentina}
\begin{abstract}
We extend the definition of non-adiabatic
entropy production given for Markovian systems in 
[M. Esposito and C. Van den Broeck, Phys. Rev. Lett. \textbf{104} 090601, (2010)],
to arbitrary non-Markov ergodic dynamics. We also introduce a notion of \emph{stability} characterizing
non-Markovianity. 
For stable non-Markovian systems, the non-adiabatic entropy production
satisfies an integral fluctuation theorem,
leading to the second law of thermodynamics for transitions between non-equilibrium steady-states.
This quantity can also be written as a sum of products of generalized fluxes and forces, thus being suitable for
thermodynamics. On the other hand, the generalized 
fluctuation-dissipation relation also holds, clarifying that the conditions for it to be satisfied 
are ergodicity and stability instead
of Markovianity. We show that in spite of being counter-intuitive, the stability criterion introduced
in this work may be violated in
non-Markovian systems even if they are ergodic, leading to a violation of the fluctuation theorem and the generalized
fluctuation-dissipation relation. Stability represents then a necessary condition for the above
properties to hold and explains why the generalized fluctuation-dissipation relation has remained elusive in the study
of non-Markov systems exhibiting non-equilibrium steady-states.
\end{abstract}

\maketitle
\section{Introduction}
\label{sec:Intro}
The lack of a formalism useful to describe, on the same footing,
 many different out of equilibrium systems, is one of the most important
unsolved problems in thermostatistics. While the equilibrium thermodynamics
for ergodic systems can be derived from the statistical theory of Gibbs, out of equilibrium
systems exhibit a great diversity, thus making difficult the unification of their description
within a single theory. 

Fluctuation theorems (FT)\cite{Evans-Cohen-Morris,Gallavotti-Cohen,Kurchan,Lebowitz,
Jarzynski,Crooks,Kurchan2,Hatano-Sasa,Seifert1,Seifert2,Harris-Schuetz} 
are exact relations for the probability distributions for the values
 $W$ of observables $\mathcal{W}$ which are functionals of the stochastic state-space system trajectories
(e.g. work, heat or more generally, different forms of trajectory-dependent entropy production) in processes
driven by arbitrary protocols. These relations, valid even beyond the linear regime, contain as particular cases
the Green-Kubo formula, the Onsager reciprocity relations and the second law of thermodynamics for systems
with equilibrium steady states in the limit of small entropy production\cite{Kurchan,GallavottiOnsa}. For systems
with non-equilibrium steady states (NESS), they provide a generalization of the second law of thermodynamics\cite{Hatano-Sasa}
which gives a formal framework for the phenomenological theory of Oono and Paniconi\cite{Oono-Paniconi}
 for the thermodynamics of NESS. In this case, also a generalized fluctuation-dissipation relation (FDR)
can be derived\cite{Richardson, Agarwal, Risken, Falcioni, Parrondo,
Maes, Verley1, Verley2, Verley3, FDRs}, as well as some `symmetry' 
relations for the response functions involved in this generalized FDR\cite{us1,us2}. For a recent review on fluctuation
theorems and their applications, see \cite{Seifert-Rev}. 

For all these reasons, FTs seem to be a good starting point in order to derive a more general theory unifying
the physics of many different out-of-equilibrium systems. In fact, many features of Markovian systems
are well described within the same formalism. 

In spite of this, non-Markovian systems demand much more work and comprehension.

In the previous recent years, some important progress has been made in this line. For example, the work fluctuations theorems
(Jarzinsky identity and Crooks fluctuation theorem) as well as the fluctuation theorem for the total entropy production
have been derived for non-Markvovian systems evolving with ergodic dynamics
\cite{Seifert-non-Markov,Ohkuma-Ohta, Mai-Dhar, BiroliFTs}. On the other hand, an asymptotic FT relevant for glassy dynamics
has been derived in \cite{Zamponi}, while the validity of the work relations has been studied in the context of anomalous dynamics
for the first time in \cite{Klages}. A discussion on the role of substitute Markov processes has been presented in
\cite{Puzzle}.

However, a lot of work remains to be done. For systems
with NESS the entropy production
associated to the total heat exchanged with the bath does not represent a relevant lower bound for the entropy change in the
 system \cite{Hatano-Sasa}. In fact, the entropy production which gives the most accurate lower bound should be related to the
transitions of the system between different steady states. When detailed balance holds, all the heat exchanged with the reservoir
is used for the system in order to jump from one steady state to the other, however, without detailed balance, part of this
heat is used in order to maintain an steady-state with non-vanishing currents. In this case, the correct lower bound is given by the
excess entropy, associated to the excess heat, which is the energy delivered or absorbed for the system when it jumps between
different steady states \cite{Hatano-Sasa}. Then, a description in terms of the Hatano-Sasa functional,
or more generally, in terms of the non-adiabatic entropy production \cite{EspositoPRL, Esposito}
is needed. As far as we know, these quantities have not
been studied yet for non-Markov dynamics. 

The aim of this paper is to fill this gap. In fact, we show that the non-adiabatic entropy production,
 as defined for Markovian systems, can satisfy
an integral fluctuation theorem (IFT) if the system fulfills certain ``stability'' condition which we discuss below. 
In this context it is possible to extend
the second law for transitions between NESS and the validity of the  generalized FDR to non-Markov dynamics.

The paper is organized as follows. In section \ref{sec:prelim} we present the basic notations and introduce the evolution operators
for some ``substitute Markov processes'' relevant for the definition of generalized fluxes.
In section \ref{sec:stability} we introduce our concept of stability and discuss an important property of stable systems. 
In section \ref{sec:NEP} we introduce the non-adiabatic entropy production and we 
derive some of its most fundamental properties, as the validity of an IFT. In this section we also discuss the extension
of the NESS thermostatistics to non-Markovian ergodic and stable systems.
In section \ref{sec:Discussion} we study a model system violating the stability criterion: a particle dragged through
a viscoelastic liquid.
The
generalized FDR is tested for an overdamped harmonic oscillator coupled to two non-Markovian
baths at different temperatures in section \ref{sec:Overdamped}. Finally, some concluding remarks are given
in \ref{sec:conclusions}. 
\section{Preliminaries}
\label{sec:prelim}
\subsection{Basic definitions and notations}
\label{subsec:def}
To start, we will consider continuous time processes for a system that can be described either by
discrete or continuous configurations. A configuration will be denoted by $x$, which may
generically represent a single variable, a vector, or a field. The system
can be additionally driven by a set of time-dependent external para\-meters which will be denoted by
$\lambda$.

Trajectories from the initial time $t_i=0$
to the final time $t_f=t$ will be denoted by vectors, 
so that $\evec=\{e(\tau)\}_{\tau=0}^{t}$, with $e=(x,\lambda)$.

Otherwise stated, stochastic functionals (like entropy productions)
will be denoted by calligraphic capi\-tal letters, while specific va\-lues
for these functionals will be denoted by the same, capital but latin character.
For example, the non-adiabatic entropy production functional will be denoted by
$\Sna[\xvec;\lvec]$, while for the specific values of this quantity we will write
$\Snal$.

We assume, as a fundamental requirement in our theo\-ry, that the system is ergodic.

Let us denote by $p(x,t)$ the probability density function (PDF)
for the system to be in the vicinity of configuration $x$ at time $t$ and by $\ps(x;\lambda)$
the steady state PDF for a constant protocol $\lambda(t)=\lambda$. Given the non-Markovian nature
of the process, one should be careful in order to introduce this distribution, as the system  
at any time (except for the initial time $t_i$)
is correlated with its previous history. Ergodicity ensures that at cons\-tant
protocols the system forgets its initial preparation for long enough times, so $\ps(x;\lambda)$ should be thought 
as the limiting PDF reached by the system at time $t\rightarrow\infty$,
 if the protocol is stopped at time $t'\ll t$ with value 
$\lambda(t')=\lambda$.

The initial PDF
will be assumed arbitrary and denoted by $\pin(x)$. On the other hand, conditional probabilities 
will be denoted by $K$.

Finally, the evolution operators for the substitute Markov processes (which will be introduced in the next
paragraph), will be generically denoted by $\mL$.
\subsection{Substitute Markov processes}
\label{subsec:Subs-Markov}
In general terms, the evolution equation for the one-time probability density of non-Markovian processes
involves convolutions in time, making explicit the fact that the system is correlated with its past. Fokker-Planck
equations with fractional time derivatives constitute very  known examples of this fact. However, the evolution
equation can, under very general conditions, be written in a convolutionless (not memoryless) 
form in terms of an evolution operator
related to some `substitute' Markov process having the same one-time probability as the origi\-nal non-Markovian process
\cite{HT,HT2}. The evolution equation reads
\begin{equation}
 \label{subs-markov}
\partial_tp(x,t)=\mL(t;t_i,[\lvec])p(x,t),
\end{equation}
where we have made explicit the fact that the evolution operator depends on the initial time, and also on the
whole protocol up to time $t$. This reflects the effect of the memory of the process. The form of the operator
also depends on the initial PDF of the system at time $t_i$, where it is prepared without any correlation with 
pre\-vious times. This means that to each particular preparation of the system, corresponds a different Markovian
substitute. When the process is Markovian, the correlation with the previous history is no longer present and
the given operator generates a semigroup, independently of the initial PDF. In this case one recovers the usual Master
(Fokker-Planck or Kramers) equation for discrete (continuous) variables.

In what follows, we will drop for simplicity the initial time and the protocol dependency on the arguments of the 
evolution operator, so we will just write $\mL(t)$.

The evolution equation (\ref{subs-markov}) can be easily derived from the general properties of conditional probabilities\cite{HT}.
In fact, let $K(x,s|x',t)$ be the conditional probability for the system to be in the vicinity of $x$ at time $s$ given that
it was around $x'$ at time $t$. One then can write for any process, being or not Markovian
\begin{equation}
 \label{cond-proper}
p(x,s)=\int dx'K(x,s|x',t)p(x',t).
\end{equation}
It is worth noting that if the process is non-Markovian, the conditional probability will depend on the whole history
up to time $s$. From the previous equation it is easy to see that
\begin{equation}
 \label{cond-proper1}
\partial_tp(x,t)=\int dx'\hM(x,x',t) p(x',t),
\end{equation}
where $\hM(x,x',t)= \lim_{s\rightarrow t^+}\partial_sK(x,s|x',t)$,
from where it is immediate to identify the operator $\mL$.
\section{Stability}
\label{sec:stability}
We say that a non-Markovian system is stable if it, being prepared at the initial time (where there is 
no correlation with the previous history) with the steady-state PDF corresponding to $\lambda$, evolves in such way that
if $\lambda$ does not change
in time, then for all times the PDF remains unchanged. 
 With a Markovian process in mind, it may seem to be counter-intuitive that the stability criterion could be violated, 
but for non-Markovian dynamics the violation may indeed occur in many realistic models. In order to understand why is it possible, 
we first make the reader to note that for Markovian systems there is no difference between preparing the system in an arbitrary state in
a very remote moment in the past and to start to observe it at $t=0$, when it has evolved to its steady-state, 
and to directly prepare the system in the steady-state at $t=0$. This is so, because previous history does not matter at all. However,
this is not the case for non-Markovian dynamics. Consider the conditional probability for the system to be around $x$ at time $t$,
 given that it was for sure at $x_0$ at time $t_0=0$, in two different contexts, when the system is prepared initially at $t=0$ with
the steady-state PDF without
any correlation with its previous history and when the system were prepared at $t=-\infty$ in an arbitrary state, letting the value
of $\lambda$ unchanged. Let us denote these propagators by $K(x,t|x_0)$ and $K_\infty(x,t|x_0)$ respectively. It is clear that, 
in general, $K(x,t|x_0)\neq K_\infty(x,t|x_0)$. Also note that $\int dx_0K_\infty(x,t|x_0)\ps(x_0;\lambda)=\ps(x;\lambda)$.
Then, it may be possible that
\begin{equation}
 \label{inequality}
\int dx_0K(x,t|x_0)\ps(x_0;\lambda)\neq \ps(x;\lambda),
\end{equation}
leading to the instability that we are describing. In words, in both cases the position at $t=0$ is sampled from the steady-state
PDF, but in the first case, the system is not really in its steady-state which, for non-Markovian systems, is not entirely
determined by the one-time PDF.

The stability criterion introduced above is in fact abstract since in practice the typical, and to be best of our knowledge
 the only, achievable way to prepare a system so that it is described by its steady PDF, is allowing it to evolve for a 
long enough time keeping
the values of the external parameters fixed, as considered in \cite{Seifert-non-Markov}. However, we will see below
that one does not actually need to be able to prepare the system sampling its initial values from the steady PDF, and that
considering an arbitrary initial distribution, the problem of the validity of the fluctuation theorem for the
non-adiabatic entropy production is mapped  to the problem of its validity in the hypothetical case we are discussing
(see equations (\ref{FTSna}) and (\ref{FTSna1}) below).
Then, the main lesson extracted from this abstract concept is that the lack of
stability gives information about the dynamics of the system in realistic situations: 
the fluctuation
theorem for the non-adiabatic entropy production does not hold for \emph{any} initial preparation, even 
in experimentally accessible conditions if one considers the evolution of the system at finite times and not
in the asymptotic regime considered, for example, in \cite{Zamponi}. A practical way to identify unstable
systems in realistic situations will be discussed in terms of the fluctuation-dissipation theorem in a future
work \cite{Rei-Viv}. 

To continue with our discussion, let us introduce the Hatano-Sasa functional $\Yhasa$ defined as \cite{Hatano-Sasa}
\begin{equation}
 \label{hs-functional}
\Yhasa[\xvec;\lvec]=\int_{0}^{t}d\tau\dot{\lambda}(\tau)\partial_\lambda\phi(x(\tau);\lambda(\tau)),
\end{equation}
with $\phi(x;\lambda)=-\ln\ps(x;\lambda)$. Let us also introduce $P_{\tt s}(x,Y,t)$, the joint PDF 
for the system to be in configuration $x$ at time $s=t$, having
observed a value of the Hatano-Sasa functional $Y$, when the initial condition is sampled from the steady state PDF, and
its Laplace transform $\hP(x,\hY,t)=\int dY e^{-Y\hY}P_{\tt s}(x,Y,t)$. Let us denote $\hP(x,t)=\hP(x,\hY=1,t)$. It turns out
that for stable systems, we have the identity
\begin{equation}
 \label{stability-def}
\hP(x,t)=\ps(x,\lambda(t)).
\end{equation}
The proof of this statement is based on a general property of \emph{correctly} defined substitute Markov evolution operators which
has been shown in \cite{Seifert-non-Markov}, say
\begin{equation}
 \label{property}
\mL(t)\ps(x;\lambda(t))=0\quad \forall t\ge 0, 
\end{equation}
for arbitrary protocols if the system is prepared in its steady-state. 
It is worth to say that, for the evolution operator to be correctly defined, a full phase-space description 
(inclusion of all the degrees of freedom) is needed,
as suggested by Ref.\cite{Puzzle}. In that reference, it has been shown that for correlated noises the velocity ceases to be
an slow variable and should be included in the determination of the substitute evolution operator, 
even if one is studying the probability
of a quantity which does not depend on velocity.
Equation (\ref{stability-def}) can be shown by simple inspection by noting the evolution equation for $\hP(x,t)$
\begin{equation}
 \label{FTSna3}
\partial_t\hP(x,t)=\mL(t)\hP(x,t)-\dot{\lambda}(t)\partial_\lambda\phi(x;\lambda(t))\hP(x,t),
\end{equation}
with initial  condition $\hP(x,0)=\ps(x;\lambda_0)$ and using the identity given by equation (\ref{property}).
Equation (\ref{FTSna3}) can be obtained by Laplace transforming the evolution equation for $P_{\tt s}(x,Y,t)$,
which corresponds to  the same equation for the PDF of the $x$ variable plus an extra term associated to the 
current in the $Y$ direction $J_Y(t)=\dot{\lambda}(t)\partial_\lambda\phi(x;\lambda(t))P_{\tt s}(x,Y,t)$
\begin{equation}
 \label{FTSna4}
\partial_tP_{\tt s}(x,Y,t)=\mL(t)P_{\tt s}(x,Y,t)-\partial_YJ_Y(t).
\end{equation}
The above reasoning seems to be correct, but there is some important condition that has to be satisfied: stability.
In fact, if stability does not hold, then (\ref{property}) and correspondingly (\ref{stability-def}) do not hold either. In order 
to see this, let us assume that (\ref{property}) holds for arbitrary protocols, but the system is unstable. Then, for
constant protocols one should have,
by virtue of instability, that $p(x,dt)\neq\ps(x;\lambda_0)$. On the other hand, by virtue of (\ref{property}) we have
$p(x,dt)=\ps(x;\lambda_0)+dt\mL(0)\ps(x;\lambda_0)=\ps(x;\lambda_0)$. This contradiction is solved only if (\ref{property})
does not hold for unstable systems.

As previously announced, in section \ref{sec:Discussion} we study a simple unstable system. 
The skeptic reader could argue that this instability is artificial, since the violation of (\ref{property}) may be associated
to a by-product of the substitute Markov process provided that one does not have a method to define it correctly in
the general case. For this reason, we discuss the instability of the model system presented in  \ref{sec:Discussion}
not by means of substitute Markov processes, but directly using generalized Langevin equations (GLE) methods, which are
always correct \cite{Puzzle}. On the other hand, in section \ref{sec:Overdamped} we study an stable system, where the
generalized FDR holds.

Before finishing with this section, we would like to briefly discuss about some systems where stability could be violated. If one
considers out-of equilibrium degrees of freedom performing non-Markovian dynamics, some ``effective'' non-local time-dependency on the 
protocol can be self-generated by performing local transformations of variables, a feature exclusively associated
to non-Markovianity (see section \ref{sec:Discussion}). On the other hand, if the non-Markovian 
noise acting on the system does not satisfy an equilibrium FDT of the
second kind (at least in terms of an effective temperature), the system is unstable.

\section{Non-adiabatic entropy production}
\label{sec:NEP}
\subsection{Definition}
\label{subsec:NEP}
 This paragraph is devoted to present the main definition of non-adiabatic entropy production
 and its physical meaning. This entropy production
have been introduced for the first time in Refs. \cite{EspositoPRL,Esposito} for Markovian systems. We will define it here
in the same way, but it will be written in a different (although equivalent) form. We define
\begin{equation}
 \label{sna}
\Sna[\xvec;\lvec]=-\ln\frac{p(x(t_f),t_f)\ps(x(t_i);\lambda(t_i))}{\pin(x(t_i))\ps(x(t_f);\lambda(t_f))}
+\Yhasa[\xvec;\lvec],
\end{equation}
 Note that, as in \cite{Esposito}, a splitting of the non-adiabatic entropy production into a boundary
contribution and a driving contribution becomes apparent. The first term, accounts for the relaxation of the system
to the steady-state, while the second term is only 
non-zero in the presence an external protocol.

It is also worth saying that in \cite{EspositoPRL} this quantity have been defined as the logratio of the path probabilities
of two different systems, obtaining (\ref{sna}) as a result. We use here (\ref{sna}) as formal definition, avoiding any
reference to path proba\-bilities and hence, making this definition ex\-tensible to non-Markov dynamics without identifying
 any `dual' dynamics, since, even if for Markovian systems the dual dynamics can be straightforwardly associated to a system with
different interactions \cite{us1,us2,Jarzynski1}, for non-Markov dynamics this identification could be more intrincate.
\subsection{Integral fluctuation theorem}
\label{subsec:IFT}
In order to start with the derivation of the IFT for the non-adiabatic entropy production, 
we point out first that, according to equation (\ref{sna}) we can
write
\begin{equation}
 \label{expsna}
e^{-\Snal}=\frac{p(x,t)\ps(x_0;\lambda_0)}{\pin(x_0)\ps(x,\lambda(t))}e^{-Y},
\end{equation}
where $x_0$ corresponds to the initial position while $\lambda_0=\lambda(0)$. Let us now introduce the conditional probability
for the system to be in configuration $x$ at time $t$, and having
observed a value of the Hatano-Sasa functional $Y$, given that it was in configuration $x_0$ at $t_i=0$. The initial value of 
the Hatano-Sasa functional has not to be specified since it is always zero. Let us denote this conditional probability by
$K(x,Y,t|x_0)$. For any observable of the form $\Lambda(x,Y,t,x_0)$ one can write
\begin{widetext}
\begin{equation}
 \label{average}
\langle\Lambda(x,Y,t,x_0)\rangle=
\int dxdx_0dY\Lambda(x,Y,t,x_0)K(x,Y,t|x_0)\pin(x_0),
\end{equation}
which implies
\begin{equation}
 \label{FTSna}
\langle\e^{-\Snal}\rangle=\int dx_0\pin(x_0)\int dx\int dY\frac{p(x,t)\ps(x_0;\lambda_0)}{\pin(x_0)\ps(x;\lambda(t))}
 K(x,Y,t|x_0)\e^{-Y}=\int dx\frac{p(x,t)}{\ps(x;\lambda(t))}\int dY P_{\tt s}(x,Y,t)\e^{-Y},
\end{equation}
 \end{widetext}
which directly leads to
\begin{equation}
 \label{FTSna1}
\langle\e^{-\Snal}\rangle=\int dx\hP(x,t)\frac{p(x,t)}{\ps(x;\lambda(t))}.
\end{equation}
If the system is stable in the sense discussed in the previous section, we have
\begin{equation}
 \label{FTSna2}
\langle\e^{-\Snal}\rangle=1.
\end{equation}

Equation (\ref{FTSna2}) is the first main result of this paper. It encodes most of the fundamental aspects for
the non-adiabatic entropy production  to be a meaningful thermodynamic quantity for ergodic systems. It is worth noting
that if the invariant measure of the system corresponds to the Boltzmann-Gibbs distribution and the system
is initially prepared in this state, the previous result
reduces to the Jarzynski relation, already derived in \cite{Seifert-non-Markov} for general non-Markovian ergodic
systems following an approach based on Markov substitute processes, and in \cite{Ohkuma-Ohta,BiroliFTs} for
non-linear generalized Langevin systems by means of an approach based on functional probabilities of trajectories.
  
\subsection{Second law}
\label{subsec:thermo-nep}
Let us introduce
the excess entropy functional $\Sex$ \cite{Hatano-Sasa} as
\begin{equation}
 \label{sex}
\Sex[\xvec;\lvec]=-\int_0^t d\tau\dot{x}\partial_x\phi(x;\lambda).
\end{equation}
For discrete spaces, the integral in (\ref{sex})
should be replaced by a sum as $\sum_{k=1}^{N}[\phi(x_{k-1};\lambda_{t_k})-\phi(x_{k};\lambda_{t_k})]$, where 
$k=1,2,\ldots,N$ labels the set of time instants when the system jumps between different configurations.
From (\ref{sna}) and (\ref{sex}) we can write
\begin{equation}
 \label{sex-sna}
\Sna=\Delta\Ss+\Sex,
\end{equation}
with $\Delta\Ss=-\ln p(x(t),t)/\pin(x_0)$ the entropy change of the system. From (\ref{FTSna2}), (\ref{sex-sna}), and the Jensen
inequality, it follows
\begin{equation}
 \label{second-law}
\langle\Delta\Ss\rangle\ge-\langle\Sex\rangle.
\end{equation}
Equation (\ref{second-law}), which is a direct consequence of (\ref{FTSna2}), is the second main result of our paper.
It represents the second
law of thermodynamics for transitions between NESS exactly as expressed in \cite{Hatano-Sasa}.                                                                                                      
\subsection{Generalized fluxes and forces}
\label{subsec:flux-forc}
Another important property (not directly derived from (\ref{FTSna2}) but also crucial in order to build
a coherent thermostatistics) is that the time derivative of the average non-adiabatic entropy production
 can be expressed as  a sum of products of generalized
fluxes and forces, as for Markov dynamics \cite{Esposito}. 
From the definition (\ref{sna}), it follows that
\begin{equation}
 \label{snadot}
\frac{d}{dt}\langle\Sna\rangle=-\int dx\partial_tp(x,t)\ln\frac{p(x,t)}{\ps(x;\lambda(t))}.
\end{equation}
Recalling now (\ref{cond-proper1}), we can write
\begin{equation}
 \label{snadot-1}
\frac{d}{dt}\langle\Sna\rangle=-\int dxdx'\hM(x,x',t)p(x',t)\ln\frac{p(x,t)}{\ps(x;\lambda(t))}.
\end{equation}
From (\ref{cond-proper1}) and the normalization condition for PDFs, one can see that $\int dx\hM(x,x',t)=0$. Then, we 
can safely add a zero to equation (\ref{snadot-1}) as
\begin{equation}
 \label{a-zero}
\int dxdx'\hM(x',x,t)p(x,t)\ln\frac{p(x,t)}{\ps(x;\lambda(t))}=0,
\end{equation}
obtaining
\begin{equation}
 \label{snadot-2}
\frac{d}{dt}\langle\Sna\rangle=-\int dxdx'\J(x,x',t)\ln\frac{p(x,t)}{\ps(x;\lambda(t))},
\end{equation}
with the fluxes $\J(x,x',t)=\hM(x,x',t)p(x',t)-
\hM(x',x,t)p(x,t)$. Note that $\J(x,x',t)=-\J(x',x,t)$, as it should be. Changing now $x$ by $x'$
in (\ref{snadot-2}), summing the resulting equation term by term with (\ref{snadot-2}) and dividing by two, 
we finally obtain
\begin{equation}
 \label{forces-fluxes}
\frac{d}{dt}\langle\Sna\rangle=-\frac{1}{2}\int dxdx'\J(x,x',t)\F(x,x',t),
\end{equation}
with the forces $\F(x,x',t)=
\ln\frac{\ps(x';\lambda(t))p(x,t)}{\ps(x;\lambda(t))p(x',t)}$. Equation
(\ref{forces-fluxes}) constitutes the third main result of our work.

For general non-Markovian dynamics it is a hard task to build the
evolution operator for the Markovian substitute process (except for Gaussian and two-level systems \cite{HT, HT2}), so,
the generalized currents can be hard to compute, however, equation (\ref{forces-fluxes})
may be very important from the conceptual (and hopefully also from the experimental) point of view. We also point out
that this property is valid even if the system is unstable.
\subsection{Generalized fluctuation-dissipation relation}
As expressed previously, if the stable system is initially prepared in the steady-state compatible with some values of
the external protocols, then equation (\ref{FTSna2}) reduces to the Hatano-Sasa identity
\begin{equation}
 \label{HS-identity}
\bigg\langle\exp\bigg[-\sum_i\int_0^td\tau\dot{\lambda}_i\partial_{\lambda_i}\phi(x;\lambda_i)\bigg]\bigg\rangle=1,
\end{equation}
where we have explicitly introduced the index $i$ to label all the external parameters.
Suppose that at the initial time we have $\lambda_i(t=0)=\lambda_{i0}$ and that for $t>0$ we have 
$\lambda_{i}(t)=\lambda_{i0}+\delta\lambda_i(t)$, with $|\delta\lambda_i/\lambda_{i0}|\ll1$.
Then, introducing the observables $b_i(t)=\partial_{\lambda_i}\phi(x(t);\lambda_0)$ and repeating
the same steps as in Ref. \cite{Parrondo}, we have  the generalized FDR
\begin{equation}
 \label{GFDR}
\langle b_i(t)\rangle=\sum_j\int_0^tdt'\frac{d}{dt}\langle b_i(t)b_j(t')\rangle_{\tt ss}\delta\lambda_j(t'),
\end{equation}
 where $\langle\ldots\rangle$ denotes averages in the perturbed system, while $\langle\ldots\rangle_{\tt ss}$ denotes
averages in the unperturbed system, where the parameters are kept fixed at their initial values. Note that, if stable, 
the system prepared in the steady-state remains there always as long as one does not perturb it.
This justifies the double-s subscript. 

Equation (\ref{GFDR}) constitutes the fourth main result of our paper. It is, in our opinion,
 a rather important result since it is commonly
claimed to hold exclusively for Markovian dynamics. This is, to the best of our knowledge, the first time that this relation
is extended, in this particular way,
 to non-Markovian systems, clarifying that the crucial conditions for (\ref{GFDR}) to hold are ergodicity and stability and 
not Markovianity.
We however point out that FDRs for non-Markov dynamics have been studied before (see for example the pioneering work \cite{HT}, and
the more recent work  \cite{Narayan}).

\section{A model system exhibiting instability}
\label{sec:Discussion}
In this section we will study a model system exhibiting instability. In this case, the Hatano-Sasa relation is
violated and correspondingly, the modified FDR does not hold.
In this model, we consider the steady-state distribution
of a genuine non-equilibrium degree of freedom for which the equation of motion is obtained from the original equation of motion
 by means of a local transformation, generating a non-local dependency on the external protocol.
The system is ergodic, however if it is prepared at $t=0$  without any correlation with the past, in such a way that
the initial position is sampled from the corresponding steady-state PDF, 
and the external protocol is kept constant, it abandons its initial state and returns
to it after a transient time.    

Consider the following GLE: 
\begin{equation}
 \label{langevin}
\int_0^td\tau\gamma(t-\tau)\dot{x}(\tau)=-k(x(t)-x_c(t))+\xi(t),
\end{equation}
where the Gaussian noise $\xi(t)$ have zero mean and a second cumulant 
$\langle\xi(t)\xi(t')\rangle=T\gamma(|t-t'|)$, with $T$ the temperature
of the bath. This relation is not sufficient in order
to ensure the ergodicity of the dynamics (see for example \cite{Hergo}). We here assume
that the system is ergodic.
We also point out that equation (\ref{langevin}) can serve as a model for a particle dragged through a
viscoelastic liquid by an optical trap, which is experimentally accessible \cite{visco}.

If we consider as the external protocol the position of the trap center, we see that the system
described by (\ref{langevin}) reaches an equilibrium state for constant $x_c$, which means that the
degree of freedom $x$ is able to equilibrate. Imagine now that we consider as external protocol not the position
of the trap center, but its velocity $v_c(t)$  as in the experiment in \cite{Trepagnier}, 
such that a constant protocol means constant velocity. In this case, the
variable $x$ is a genuine out-of-equilibrium degree of freedom which is not even allowed to reach an steady value.
We can however recover a variable capable to reach an steady value if we consider the quantity $y(t)=x(t)-x_c(t)$, which
is also a genuine out of equilibrium degree of freedom. 
In terms
of the variable $y$, the equation of motion reads
\begin{equation}
 \label{langevin1}
\int_0^td\tau\gamma(t-\tau)\dot{y}(\tau)=-ky(t)+F(t)+\xi(t),
\end{equation}
where the force $F$ is given by
\begin{equation}
 \label{force}
F(t)=-\int_0^td\tau\gamma(t-\tau)v_c(\tau).
\end{equation}
Note that even for constant velocity, this force is time dependent. This non local in time dependency on
the external protocol will have drastic consequences. Also note that for Markov dynamics with $\gamma(t)\sim\delta(t)$,
the dependency on the external protocol becomes local. 

The solution of (\ref{langevin1}) can be found by means of the Laplace transform and reads
\begin{equation}
 \label{solution}
y(t)=G_1(t)y_0-\int_0^td\tau G_1(t-\tau)v_c(\tau)+\eta(t),
\end{equation}
where $\eta(t)=\int_0^td\tau G_2(t-\tau)\xi(\tau)$. The quantities $G_1(t)$ and $G_2(t)$ are the inverse Laplace
transforms of $\hat{G}_1(u)$ and $\hat{G}_2(u)$ given by
\begin{equation}
 \label{transforms}
\hat{G}_2(u)=[u\hat{\gamma}(u)+k]^{-1}; \quad \hat{G}_1(u)=\hat{\gamma}(u)\hat{G}_2(u).
\end{equation}
The noise $\eta$ has zero mean and correlator
\begin{equation}
 \label{noise-corr}
\Delta(t,t')=\langle\eta(t)\eta(t')\rangle=\frac{T}{k}\big[G_1(|t-t'|)-G_1(t)G_1(t')\big],
\end{equation}
as can be easily shown by direct calculation in the Laplace space. Note that ergodicity requires
that $\lim_{t\rightarrow\infty} G_1(t)=0$, while compatibility with the initial conditions demands
$G_1(t=0)=1$. We also note that from (\ref{transforms}) we can obtain the following identity
\begin{equation}
 \label{G1H}
G_1(t)=1-kH(t),
\end{equation}
with $H(t)=\int_0^td\tau G_2(\tau)$. One can see, by using (\ref{solution}), (\ref{noise-corr}), and (\ref{G1H}),
and using the properties of $G_1$
that for constant $v_c$ the system reaches the following distribution
\begin{equation}
 \label{steady1}
\ps(y;v_c)=\sqrt{\frac{k}{2\pi T}}\exp\bigg[-\frac{k(y+\gef^\infty v_c)^2}{2T}\bigg],
\end{equation}
where $\gef^\infty=\gef(t\rightarrow\infty)$, and $\gef(t)=\int_0^td\tau G_1(\tau)$.
Now let us assume that the initial position of the particle is sampled from this distribution, which means that
\begin{align}
 \langle y_0\rangle=-\gef^\infty v_c,\\
 \langle\delta y_0^2\rangle=\frac{T}{k},
\end{align}
where $\delta y_0=y_0-\langle y_0\rangle$. Using this in (\ref{solution}) 
for constant $v_c$, we obtain that still $\langle\delta y^2(t)\rangle
=\langle\delta y_0^2\rangle=\frac{T}{k}$, with $\delta y(t)=y(t)-\langle y(t)\rangle$, however, the mean value of the process is
\begin{equation}
 \label{dif-mean}
\langle y(t)\rangle=-\big[\gef^\infty G_1(t)+\gef(t)\big]v_c\neq\langle y_0\rangle.
\end{equation}
As the process is Gaussian, this is enough to ensure that $p(y,t)\neq\ps(y;v_c)$ for finite times, 
thus, the system is
unstable.
Note however that 
\begin{equation}
\lim_{t\rightarrow\infty}\langle y(t)\rangle=\langle y_0\rangle, 
\end{equation}
which means that the system decays again to the steady-state PDF. 

We remark once again that this instability is the result of sampling the initial position from the steady-state
PDF at the very beginning of the evolution, where the system is uncorrelated with the bath and its previous history. Once
the system reaches the steady state after a long time in interaction with the thermal bath, this instability disappears. 
In other words, in this model the instability is associated to the fact that initially, even if the system is prepared
with the steady-state PDF, the non-local force defined by (\ref{force}) depends on time even for constant protocols. This
does not happen for a system prepared far away in the past. When this time-depending force relaxes, the instability disappears. 

Finally, it is worth noting that in the Markovian case the instability is no longer present. In fact, if $\gamma(t)=\delta(t)$,
then $G_1(t)=e^{-kt}$ and $\gef(t)=\gef^\infty [1-G_1(t)]$, with $\gef^\infty=1/k$. Using this in (\ref{dif-mean}), we immediately
see that $\langle y(t)\rangle=\langle y_0\rangle$ for all times.

\section{Overdamped harmonic oscillator coupled to two non-Markovian baths}
\label{sec:Overdamped}
We will now test the validity of (\ref{GFDR}) for non-Markovian ergodic systems considering a very simplistic
(and more unrealistic than the previous one) model, but useful in order to illustrate our findings. Consider
an overdamped harmonic oscillator, coupled to two non-Markovian baths
\begin{align}
 \label{lang-oscillator}
\int_0^td\tau[\gamma_1(t-\tau)+\gamma_2(t-\tau)]\dot{x}(\tau)=\nonumber\\
-kx(t)+f(t)+\xi_1(t)+\xi_2(t),
\end{align}
with Gaussian noises with variances $\langle\xi_\nu(t)\xi_\nu(t')\rangle=T_\nu\gamma_\nu(|t-t'|)$, $\nu=1,2$.
The force $f$ will be considered as an external protocol. We can identify the heat  exchanged with each reservoir
\begin{equation}
 \label{heat}
dQ_\nu(t)=\bigg[\int_0^td\tau\gamma_\nu(t-\tau)\dot{x}(\tau)-\xi_\nu(t)\bigg]\dot{x}(t)dt,
\end{equation}
the energy change
\begin{equation}
 \label{energy}
dE(t)=kx(t)\dot{x}(t)dt,
\end{equation}
and the work
\begin{equation}
 \label{work}
dW(t)=f(t)\dot{x}(t)dt,
\end{equation}
from where the first law of thermodynamics follows
\begin{equation}
 \label{1rst-law}
dE(t)=dW(t)-dQ(t),
\end{equation}
with the total heat $dQ=dQ_1+dQ_2$. In order to ensure the ergodicity and the stability of the dynamics, 
we take the two baths to be identical, so $\gamma_1(t)=\gamma_2(t)=\frac{1}{2}\gamma(t)$. Introducing then
the effective temperature $\Tef=\frac{1}{2}(T_1+T_2)$, we can write for the dynamics of the system
\begin{equation}
 \label{lang-oscillator1}
\int_0^td\tau\gamma(t-\tau)\dot{x}(\tau)=-kx(t)+f(t)+\xi(t),
\end{equation}
with $\langle\xi(t)\xi(t')\rangle=\Tef\gamma(|t-t'|)$. The solution of this equation is given by
\begin{equation}
 \label{solution-lang-oscillator}
x(t)=G_1(t)x_0+\int_0^td\tau G_2(t-\tau)f(\tau)+\eta(t),
\end{equation}
with $G_1$, $G_2$, and $\eta$ as given in \ref{sec:Discussion}. Then, the steady-state PDF for this
system is given by the Boltzmann-Gibbs distribution
\begin{equation}
 \label{steady-lang-oscillator}
\ps(x;f)=\sqrt{\frac{k}{2\pi\Tef}}\exp\bigg[-\frac{(kx-f)^2}{2k\Tef}\bigg].
\end{equation}
As (\ref{steady-lang-oscillator}) is similar to an equilibrium PDF, one should be tempted to believe that the average total 
entropy production rate in the steady state is zero, however, this is incorrect. One can, for example, read
an illuminating discussion about this point in Ref. \cite{Tome}, where the entropy production of a spin model
in one and two dimensions have been studied. As the authors correctly pointed out, an steady-state PDF of the
Boltzmann-like type is not sufficient to ensure that the system is in equilibrium. What really defines equilibrium
is detailed balance, or in a more macroscopic language, the vanishing of the average entropy production.
In this sense, one should also note that, even if
the coarse-grained description given by (\ref{lang-oscillator1}) correctly estimates the non-adiabatic entropy production,
it severely underestimates the total entropy production, as pointed out in the second reference in \cite{Esposito} for a Markovian
system. Then,
(\ref{steady-lang-oscillator}) represents a genuine NESS.

Let us assume that the system is initially prepared in the steady state associated to the force $f_0$ and that
$f(t)=f_0+\delta f(t)$. In this case we identify the observable $b(t)$ as
\begin{equation}
 \label{bent}
b(t)=\frac{f_0-kx(t)}{k\Tef}.
\end{equation}
Note now that we can rewrite (\ref{solution-lang-oscillator}) as follows
\begin{equation}
 \label{re-write}
x(t)=G_1(t)x_0+\frac{f_0}{k}\big(1-G_1(t)\big)+\int_0^td\tau G_2(t-\tau)\delta f(\tau)+\eta(t),
\end{equation}
where we have made use of (\ref{G1H}). From (\ref{steady-lang-oscillator}) we see that 
$\langle x_0\rangle=f_0/k$, which implies
\begin{equation}
 \label{mean-FDR}
\langle b(t)\rangle=-(\Tef)^{-1}\int_0^td\tau G_2(t-\tau)\delta f(\tau).
\end{equation}
Now, for the unperturbed system we can write
\begin{equation}
 \label{bent-unperturbed}
b(t)=-(\Tef)^{-1}\bigg[G_1(t)\big(x_0-\frac{f_0}{k}\big)+\eta(t)\bigg],
\end{equation}
from where it follows recalling (\ref{noise-corr})
\begin{equation}
 \label{bent-corr}
\langle b(t)b(t')\rangle_{\tt ss}=(k\Tef)^{-1}G_1(t-t'); \quad t>t'.
\end{equation}
Now, using (\ref{G1H}), we can write
\begin{equation}
 \label{time-der}
\frac{d}{dt}\langle b(t)b(t')\rangle_{\tt ss}=-(\Tef)^{-1}G_2(t-t').
\end{equation}
Then, comparing (\ref{time-der}) with (\ref{mean-FDR}), we conclude that
\begin{equation}
 \label{FDR-final}
\langle b(t)\rangle=\int_0^tdt'\frac{d}{dt}\langle b(t)b(t')\rangle_{\tt ss} \delta f(t'),
\end{equation}
which completes the proof. Then, we have checked the general result (\ref{GFDR}) for a genuine non-Markovian
system with NESS.
\section{Concluding remarks and perspectives}
We have shown that, in strong contrast with Markovian systems, non-Markov dynamics may be  unstable, in the sense
that a system prepared in such a way that the initial positions are sampled from the steady-state PDF, may depart from this state
at finite times even if the external protocols are kept constant.

For stable systems the non-adiabatic entropy production satisfies an integral fluctuation theorem 
and the second law of thermodynamics for transitions
between NESS holds, exactly as for Markov dynamics. On the other hand, the generalized FDR is also verified, clarifying that, contrary
to what is often asserted that this relation only holds for Markovian systems, the conditions which need to be fulfilled are 
ergodicity and stability. However, if the stability condition fails, it turns out that the generalized FDR does not hold anymore.
We believe that this is the reason why the generalized FDR has remained elusive up to now for non-Markov dynamics. It is common to see
this issue discussed in the literature by the study of models violating the stability condition (see for example the model
discussed in Ref. \cite{Parrondo} related to a molecular motor with an internal relaxation time, which is a reliable model
for the experimental situation presented in Ref. \cite{mol-motor}).

The time derivative of the average non-adiabatic entropy production 
can be written as a sum of products of generalized fluxes and forces, even without stability. This could
be relevant for the experimental determination of this quantity if one is able to determine the steady-state
distribution of the system, since the particle current can also be determined in the experiment. On the other
hand, the determination of the currents may give direct information about the properties of the evolution
operator.

Some interesting open questions remain to be answered. First, the integral fluctuation theorem (\ref{FTSna2}) 
suggests that a detailed fluctuation theorem may
also hold if one introduces a dual system so that the non-adiabatic entropy production 
can be expressed as the logratio of the forward path probability
of the original system and the time-reversed path probability of the dual system. This also could make easy to identify
an adiabatic entropy production, such that the total entropy production can be split as in \cite{EspositoPRL}.
Second, it would be interesting to generalize the FDR to the case when the system is unstable. In
this case, we speculate that some `violation' terms should appear and it would be interesting to investigate their precise form 
and physical meaning. It would be also interesting to relate that case to the recent results presented in 
\cite{Bohec}. 
Third, some immediate improvements of the present theory can be developed in order to describe a wider variety of 
systems.
For example, an extension of this theory to describe also non-ergodic systems, can be attempted
in the spirit of Refs. \cite{infinite2ndlawlike, Verley3}. In those references, 
generalized Hatano-Sasa identities have been obtained in terms of functions
which are not related to the steady-state PDF. This may be relevant if the steady state is not univocally
determined, as it is the case for non-ergodic systems. With this improvement, the same formalism could be used in order to describe
such complex systems as glasses far from asymptotic states.

\label{sec:conclusions}

\begin{acknowledgments}
This work was supported by CNEA, CONICET (PIP11220090100051), ANPCYT (PICT2007886). We are indebted to V. Lecomte
for a criti\-cal reading of the manuscript and for pointing out to us Ref. \cite{Bohec}. We also thank
D. Dom\'inguez, A. B. Kolton and S. Bustingorry for their useful suggestions. 
\end{acknowledgments}

\end{document}